\documentclass[prd,aps,showpacs,onecolumn]{revtex4}
\usepackage{amsfonts}

\usepackage{graphicx}
\usepackage{bm}

\begin{document}

\title{Recursive Diagonalization of Quantum Hamiltonians to all order in $\hbar$}
\author{Pierre Gosselin$^{1}$, Jocelyn Hanssen$^{2}$ and Herv\'{e} Mohrbach$^{2}$}

\address{$^1$ Universit\'e Grenoble I, Institut Fourier, UMR
5582 CNRS-UJF, UFR de Math\'ematiques, BP74, 38402 Saint Martin
d'H\`eres, Cedex, France \\
$^2$ Universit\'e Paul Verlaine, Institut de Physique, ICPMB1-FR
CNRS 2843, Laboratoire de Physique Mol\'eculaire et des
Collisions, 57078 Metz, France}

\begin{abstract}
We present a diagonalization method for generic matrix valued Hamiltonians
based on a formal expansion in power of $\hbar $. Considering $\hbar $ as a
running parameter, a differential equation connecting two diagonalization
processes for two very close values of $\hbar $ is derived. The integration
of this differential equation allows the recursive determination of the
series expansion in powers of $\hbar $ for the diagonalized Hamiltonian.
This approach results in effective Hamiltonians with Berry phase corrections
of higher order in $\hbar $, and deepens previous works on the semiclassical
diagonalization of quantum Hamiltonians which led notably to the discovery
of the intrinsic spin Hall effect. As physical applications we consider
spinning massless particles in isotropic inhomogeneous media and show that
both the energy and the velocity get quantum corrections of order $\hbar
^{2} $. We also derive formally to all order in $\hbar $ the energy spectrum
and the equations of motion of Bloch electrons in an external electric field.
\end{abstract}

\maketitle

\section{Introduction}

Recently, a lot of works have shown the relevance of the Berry phase in
semiclassical Physics. For instance, a new set of semiclassical equations of
motion including a Berry phase correction was derived within a Lagrangian
formalism to account for the adiabatic wave-packet evolution of an electron
in an electromagnetic field \cite{NIU1} (see also \cite{SHINDOU}). A similar
semiclassical Lagrangian approach for a light wave-packet in a inhomogeneous
medium predicted an optical Hall effect \cite{MURAKAMI2}. However this
Lagrangian formalism, although adapted to the evolution of a wave-packet, is
only semi-classical by construction so that higher order quantum corrections
are out of reach. Moreover, it was found that the subsequent derivation of
the Hamiltonian presents some difficulties due to the presence of
Berry-phase terms \cite{NIU2}.

The second difficulty has been solved by a diagonalization procedure with
accuracy $\hbar $ for a generic matrix valued Hamiltonian \cite
{PIERREGENERAL}. This method results in an effective semiclassical diagonal
Hamiltonian with Berry phase corrections as well as noncommutative covariant
coordinates and momentum operators\textbf{.} The resulting generic equations
of motion are also corrected by Berry phase terms. In the case of a Bloch
electron in electromagnetic field \cite{PIERREGENERAL}\cite{PIERREBLOCH},
this approach leads to the same equations of motion and the same
magnetization as in \cite{NIU1}. Therefore, the semiclassical
diagonalization of the quantum Hamiltonian not only shows that there is no
trouble with the Hamiltonian in the presence of Berry phase terms, but also
presents a practical short-cut to derive an effective Hamiltonian,
noncommutative coordinates, momentum operators and equations of motion with
Berry phase corrections. Moreover, this formalism, which in our opinion,
better reflects the physical origin of the phenomena under consideration, is
a general one that has been and could still be applied to several other
systems in condensed matter or particle physics.

For instance, for spinning particles, this approach led to the discovery of
the intrinsic spin Hall effect or topological spin transport. Indeed it was
already known that Berry phases, which are in fact spin-orbit couplings,
modify semiclassical dynamics of spinning particles in electric \cite{ALAIN}
and magnetic field \cite{BLIOKH1} as well as in semiconductor \cite
{MURAKAMI1}. In this last case, the Berry phase in momentum space might be
responsible for a transverse dissipationless spin-current in the presence of
an electric field. In addition to that, spin-orbit contributions to the
propagation of light in isotropic inhomogeneous media has been the focus of
several other works \cite{ALAIN}\cite{BLIOKH2} and has led to a
generalization of geometric optics called geometric spinoptics \cite
{HORVATHY1}.

The diagonalization method of \cite{PIERREGENERAL} allows to extend these
studies to spinning particles interacting with static gravitational fields
and goes beyond previous approaches \cite{OBUKHOV}\cite{SILENKO}. It led to
the discovery of new couplings between the spin and magneto-torsion fields
which could reveal an hypothetical torsion of space \cite{PIERREPHOTON}\cite
{PIERREELECTRON}.

Therefore, the diagonalization procedure of general quantum Hamiltonian has
the advantage, compared to the semiclassical wave-packet Lagrangian
approach, to gather several different systems into one general scheme. We
find particularly interesting that our approach also gives to photons and
electrons an equal footing. Indeed the spin Hall effect which offers
promising applications both in spintronics and in spinoptics, has its
origin, within our formalism, in the noncommutativity of the coordinates for
both particles. From this point of view it is always legitimate to wonder
whether an electronic phenomenon has its photonic counterpart.

However, the diagonalization procedure mentioned above, despite its
advantages, is limited to the semi classical level. It is thus legitimate to
wonder about the possibility to go one step further, i.e. to find higher
order quantum corrections. The goal of the present article is to show how to
go beyond the semiclassical approximation, a task beyond the reach of the
Lagrangian formalism mentioned previously. The method we develop here is
entirely new and different from the semiclassical diagonalization of \cite
{PIERREGENERAL}, although based on it. It results in an effective diagonal
Hamiltonian with Berry phase corrections as well as noncommutative
coordinates and momentum operators (in the adiabatic approximation) all
written as series expansions in $\hbar $.

More precisely, we propose a procedure to diagonalize recursively in power
series of $\hbar $ a generic matrix valued Hamiltonian. The key element of
this method relies on considering $\hbar $ as a running parameter. It allows
to derive a differential equation connecting two diagonalization processes
for two very close values of $\hbar $.The recursive diagonalization can then
be obtained by integrating this differential equation order by order in $%
\hbar $. While this method does not allow an exact derivation of the
diagonalized Hamiltonian, it has nevertheless the advantage to make formally
possible the computation of quantum corrections of order higher than $\hbar $
(at least second order). To show this point and explain practically our
technique, we study as a first example, a photon in an isotropic
inhomogeneous medium at the second order in $\hbar $, giving thus
corrections to the semi classical equation we derived in \cite{PIERREPHOTON}%
. An important consequence of this computation is that the light velocity
and the photon energy get a quantum correction of order $\hbar ^{2}$. As a
second different physical application, we also provide, formally to all
order in $\hbar $, the energy spectrum and the equations of motion for Bloch
electrons in a constant external electric field. We show that the equations
of motion are formally the same to all order in $\hbar $, and that only the
Berry connections get contributions in this expansion.

The paper is organized as follows. In the next section we develop our
formalism in the case of a generic matrix valued Hamiltonian. We derive the
recursive differential equation in $\hbar $ for the diagonalized
Hamiltonian. In section 3 we discuss the semiclassical approximation and
derive the generic equations of motion with Berry phases corrections.
Section 4 is devoted to exemplify the method beyond the semiclassical level
for physically relevant systems. Section 5 is for the conclusion.

\section{ Recursive diagonalization of quantum Hamiltonian}

In this section we consider a quantum mechanical system whose state space is
a tensor product $L^{2}\left( \mathcal{R}^{3}\right) \otimes V$\ with $V$\
some internal space. In other words, the Hamiltonian of this system can be
written as a matrix $H_{0}\left( \mathbf{P,R}\right) $\ of size $\dim V$\
whose elements are operators depending on a couple of canonical variables $P$%
\ and $R$, the archetype example beeing the Dirac Hamiltonian with $V=C^{4}$%
. Our goal is to describe a diagonalization process for this matrix valued
quantum Hamiltonian $H_{0}\left( \mathbf{P,R}\right) $\ recursively as a
series expansion in powers of $\hbar $. This expansion in powers of $\hbar $%
\ gives the quantum corrections to the diagonalized Hamiltonian with respect
to the classical situation $\hbar =0$. For example, the first order
correction in $\hbar $\ corresponds to the semiclassical approximation. Let
us stress at this point that, by diagonalization, we always mean here a
unitary transformation setting the Hamiltonian in a diagonal matrix form,
the diagonal elements beeing operators depending on $P$\ and $R$. That is,
we do not aim at finding the eigenvalues, but rather to derive the diagonal
representation of Hamiltonians, that are usually relevant for the dynamics.

We will derive the $\hbar $\ expansion recursively in the following way. The
Planck constant $\hbar $\ is formally promoted to a dynamical parameter $%
\alpha $\ in order to establish a differential equation connecting the two
diagonalized Hamiltonians at $\hbar =\alpha $\ and $\hbar =\alpha +d\alpha $%
. The integration of this differential equation allows then the recursive
determination of the different terms in the expansion of the diagonalized
Hamiltonian in powers of $\alpha $.

\subsection{Set up}

To start with, consider a quantum mechanical system with canonical variables
$\mathbf{P}$ and $\mathbf{R}$, where $\mathbf{P}$ is the generator of \
translations. It could be the usual momentum, or the magnetic translation
operator of electrons in magnetic Bloch bands in solid state physics as
discussed in the section devoted to the physical applications (see also \cite
{PIERREBLOCH}).

Let consider the commutator between the canonical variables as a dynamical
parameter $\alpha $, that is
\begin{equation}
\left[ \mathbf{P,R}\right] =-i\alpha
\end{equation}

Consider now the three following assumptions :

$1$. The Hamiltonian of the system can be written as a matrix of a certain
size $H_{0}\left( \mathbf{P,R}\right) $, that is a matrix whose coefficients
depend on $\mathbf{P}$ and $\mathbf{R}$. The typical example is the free
Dirac Hamiltonian (a $4\times 4$ matrix), but as shown in \cite
{PIERREGENERAL} an electron in a periodic potential also fits in this set up.

$2$. Assume moreover, that for each value of $\alpha $, $H_{0}\left( \mathbf{%
P,R}\right) $ is exactly diagonalized through a matrix \ $U_{\alpha }\mathbf{%
(P,R})$, i.e.
\begin{equation}
U_{\alpha }\left( \mathbf{P},\mathbf{R}\right) H_{0}\left( \mathbf{P,R}%
\right) U_{\alpha }^{+}\left( \mathbf{P,R}\right) =\varepsilon _{\alpha
}\left( \mathbf{P,R}\right) \text{ if }\left[ \mathbf{P,R}\right] =-i\alpha
\label{CANONIC1}
\end{equation}
where $\varepsilon _{\alpha }\left( \mathbf{P,R}\right) $ is a diagonal
matrix.

The index in $U_{\alpha }\mathbf{(P,R})$ underlines the fact that the
diagonalization matrix depends a-priori on $\alpha $ since the
diagonalization process involves recombination of powers of $\mathbf{P}$ and
$\mathbf{R}$, and thus involves commutators of these variables (see \cite
{PIERREGENERAL}).

$3$. The diagonalization is known when $\alpha =0$, that is when $\mathbf{P}$
and $\mathbf{R}$ commute. This assumption is crucial, since it allow to
start a recursive diagonalization process in powers of $\alpha $. This last
assumption has also the advantage to be quite reasonable, since in many
cases, the diagonalization is very easy to perform when $\left[ \mathbf{P,R}%
\right] =0$. Actually, in this case the Hamiltonian can be seen as depending
only on $\mathbf{P}$ for example, $\mathbf{R}$ being an external parameter.
More about this point can be found in \cite{PIERREGENERAL}.

\subsection{The differential equation}

\subsubsection{$\protect\alpha $ as a parameter}

We will now consider $\alpha $ as a running parameter in order to find a
relation between $\varepsilon _{\alpha }\left( \mathbf{P,R}\right) $ and $%
\varepsilon _{\alpha +d\alpha }\left( \mathbf{P,R}\right) $. But to
facilitate the computation of $\varepsilon _{\alpha }\left( \mathbf{P,R}%
\right) $ as explained in \cite{PIERREGENERAL}, we first need to write the
Hamiltonian as well as all expressions involving the canonical operators $%
\mathbf{R}$ and $\mathbf{P}$, in a symmetrized form. By this, we mean that
whatever the prescription for the arrangement of the canonical variables
(full symmetrization for instance or Weyl prescription) in the initial
Hamiltonian, one will always rewrite it (and other expressions too) in a
form where all the powers of $\mathbf{P}$ have been put half on the left and
half of the right of the expression. Of course this symmetrization will
introduce terms of order $\alpha $ due to the commutation relations. Note
again that this is just a convention which facilitates the diagonalization.

To start with, consider the diagonalization at the scale $\alpha $
\begin{equation}
U_{\alpha }\left( \mathbf{P},\mathbf{R}\right) H_{0}\left( \mathbf{P,R}%
\right) U_{\alpha }^{+}\left( \mathbf{P,R}\right) =\varepsilon _{\alpha
}\left( \mathbf{P,R}\right) \text{ if }\left[ \mathbf{P,R}\right] =-i\alpha
\end{equation}
and similarly for $\alpha +d\alpha $.
\begin{equation}
U_{\alpha +d\alpha }\left( \mathbf{P},\mathbf{R}\right) H_{0}\left( \mathbf{%
P,R}\right) U_{\alpha +d\alpha }^{+}\left( \mathbf{P,R}\right) =\varepsilon
_{\alpha +d\alpha }\left( \mathbf{P,R}\right) \text{ if }\left[ \mathbf{P,R}%
\right] =-i\left( \alpha +d\alpha \right)
\end{equation}
Let us develop this last relation to the first order in $d\alpha $,
\[
\varepsilon _{\alpha +d\alpha }\left( \mathbf{P,R}\right) =U_{\alpha
}H_{0}U_{\alpha }^{+}+d\alpha \left( \partial _{\alpha }U_{\alpha
}H_{0}U_{\alpha }^{+}+U_{\alpha }H_{0}\partial _{\alpha }U_{\alpha
}^{+}\right)
\]
where for clarity we have omitted the $\mathbf{P}$ and $\mathbf{R}$
dependence in the right hand side (note that here we assume the relation $%
\left[ \mathbf{P,R}\right] =-i\left( \alpha +d\alpha \right) $ and not $%
\left[ \mathbf{P,R}\right] =-i\alpha $). This latter expression at the first
order in $d\alpha $, reduces to
\begin{eqnarray*}
\varepsilon _{\alpha +d\alpha }\left( \mathbf{P,R}\right) &=&U_{\alpha
}H_{0}U_{\alpha }^{+}+d\alpha \left( \partial _{\alpha }U_{\alpha }U_{\alpha
}^{+}\varepsilon _{\alpha }\left( \mathbf{P,R}\right) +\varepsilon _{\alpha
}\left( \mathbf{P,R}\right) U_{\alpha }\partial _{\alpha }U_{\alpha
}^{+}\right) \\
&=&U_{\alpha }H_{0}U_{\alpha }^{+}+\varepsilon _{\alpha }\left( \mathbf{P,R}%
\right) \left( \partial _{\alpha }U_{\alpha }U_{\alpha }^{+}+U_{\alpha
}\partial _{\alpha }U_{\alpha }^{+}\right) d\alpha +\left[ \partial _{\alpha
}U_{\alpha }U_{\alpha }^{+},\varepsilon _{\alpha }\left( \mathbf{P,R}\right) %
\right] d\alpha
\end{eqnarray*}
This differential equation is the main point of this paper. We now show how
to compute explicitly each term of the right hand side, to make it tractable.

\subsubsection{Computation of $\partial _{\protect\alpha }U_{\protect\alpha
}\left( \mathbf{P},\mathbf{R}\right) U_{\protect\alpha }^{+}\left( \mathbf{%
P,R}\right) +U_{\protect\alpha }\left( \mathbf{P},\mathbf{R}\right) \partial
_{\protect\alpha }U_{\protect\alpha }^{+}\left( \mathbf{P},\mathbf{R}\right)
$}

From the equality
\begin{equation}
U_{\alpha +d\alpha }(\mathbf{P},\mathbf{R)}U_{\alpha +d\alpha }^{+}(\mathbf{%
P,R)=}1\text{ when }\left[ \mathbf{P,R}\right] =-i\left( \alpha +d\alpha
\right)
\end{equation}
one obtains
\begin{equation}
d\alpha \left( \partial _{\alpha }U_{\alpha }\left( \mathbf{P},\mathbf{R}%
\right) U_{\alpha }^{+}\left( \mathbf{P,R}\right) +U_{\alpha }\left( \mathbf{%
P},\mathbf{R}\right) \partial _{\alpha }U_{\alpha }^{+}\left( \mathbf{P},%
\mathbf{R}\right) \right) =1-U_{\alpha }(\mathbf{P},\mathbf{R)}U_{\alpha
}^{+}(\mathbf{P,R)}  \label{unmoinsUU}
\end{equation}
Note that here $U_{\alpha }(\mathbf{P},\mathbf{R)}U_{\alpha }^{+}(\mathbf{%
P,R)\neq }1$, since $\left[ \mathbf{P,R}\right] =-i\left( \alpha +d\alpha
\right) $. As a consequence, we can ultimately rewrite :
\begin{equation}
\varepsilon _{\alpha +d\alpha }\left( \mathbf{P,R}\right) =U_{\alpha
}H_{0}U_{\alpha }^{+}+\varepsilon _{\alpha }\left( \mathbf{P,R}\right) \left[
1-U_{\alpha }U_{\alpha }^{+}\right] +\left[ \partial _{\alpha }U_{\alpha
}U_{\alpha }^{+},\varepsilon _{\alpha }\left( \mathbf{P,R}\right) \right]
d\alpha  \label{equadiff}
\end{equation}
we now need to rewrite $U_{\alpha }H_{0}U_{\alpha }^{+}$.

\subsubsection{Computation of $U_{\protect\alpha }\left( \mathbf{P},\mathbf{R%
}\right) H_{0}\left( \mathbf{P,R}\right) U_{\protect\alpha }^{+}\left(
\mathbf{P,R}\right) $ and final form for the differential equation.}

Here, the important point to keep in mind is that $\left[ \mathbf{P,R}\right]
=-i\left( \alpha +d\alpha \right) $, so that
\begin{equation}
U_{\alpha }(\mathbf{P},\mathbf{R)}H_{0}\left( \mathbf{P,R}\right) U_{\alpha
}^{+}(\mathbf{P,R)}\neq \varepsilon _{\alpha }\left( \mathbf{P,R}\right)
\end{equation}
To compute $U_{\alpha }H_{0}U_{\alpha }^{+}$ we introduce two ''fictitious''
variables ,$\mathbf{r,p}$, commuting with $\mathbf{P}$ and $\mathbf{R}$ and
such that $\left[ \mathbf{p,r}\right] =id\alpha $. \ We can consider $%
\mathbf{r}$ and\textbf{\ }$\mathbf{p}$ as an arbitrary couple of ''small''
canonical variables each of magnitude $\sqrt{d\alpha }$. As a consequence
\begin{equation}
\left[ \mathbf{P+p,R+r}\right] =-i\alpha
\end{equation}
so that in virtu of Eq.\ref{CANONIC1} we have the equality
\begin{equation}
U_{\alpha }(\mathbf{P+p},\mathbf{R+r)}H_{0}\left( \mathbf{P+p,R+r}\right)
U_{\alpha }^{+}(\mathbf{P+p,R+r)}=\varepsilon _{\alpha }\left( \mathbf{%
P+p,R+r}\right)  \label{UUepsilon}
\end{equation}
since $\mathbf{P+p}$ and $\mathbf{R+r}$ form a couple of canonical variables
with commutator $-i\alpha $. By the same trick, and for practical purpose,
we can write :
\begin{equation}
U_{\alpha }(\mathbf{P+p},\mathbf{R+r)}U_{\alpha }^{+}(\mathbf{P+p,R+r)=}1
\label{UUplus}
\end{equation}
As a consequence, using Eqs. \ref{UUepsilon} and \ref{UUplus} we can rewrite
our differential equation Eq. \ref{equadiff} as :
\begin{eqnarray*}
\varepsilon _{\alpha +d\alpha }\left( \mathbf{P,R}\right) &=&\varepsilon
_{\alpha }\left( \mathbf{P+p,R+r}\right) \\
&&-U_{\alpha }(\mathbf{P+p},\mathbf{R+r)}H_{0}\left( \mathbf{P+p,R+r}\right)
U_{\alpha }^{+}(\mathbf{P+p,R+r)-}U_{\alpha }(\mathbf{P},\mathbf{R)}%
H_{0}\left( \mathbf{P,R}\right) U_{\alpha }^{+}(\mathbf{P,R)} \\
&&+\varepsilon _{\alpha }\left( \mathbf{P,R}\right) \left[ U_{\alpha }(%
\mathbf{P+p},\mathbf{R+r)}U_{\alpha }^{+}(\mathbf{P+p,R+r)}-U_{\alpha }(%
\mathbf{P},\mathbf{R)}U_{\alpha }^{+}(\mathbf{P,R)}\right] \\
&&+\left[ \partial _{\alpha }U_{\alpha }\left( \mathbf{P},\mathbf{R}\right)
U_{\alpha }^{+}\left( \mathbf{P,R}\right) ,\varepsilon _{\alpha }\left(
\mathbf{P,R}\right) \right] d\alpha
\end{eqnarray*}
Now, we expand the R.H.S to the second order\ in powers of $\mathbf{p}$ and $%
\mathbf{r}$. Since those two variables satisfy non trivial commutation
relations, we choose to expand the R.H.S. on the basis $%
p_{i},r_{i},p_{i}p_{j},r_{i}r_{j},p_{i}r_{j}$ for $i\neq j,$ and $\frac{%
p_{i}r_{i}+r_{i}p_{i}}{2}$ (Birkhoff-Witt factorization theorem). As a
consequence, the term $p_{i}r_{i}$ or $r_{i}p_{i}$ have to be rearranged in
combination of $\frac{p_{i}r_{i}+r_{i}p_{i}}{2}$ and $\frac{%
p_{i}r_{i}-r_{i}p_{i}}{2}=\frac{id\alpha }{2}$.

The expansion computation is similar to the one presented in \cite
{PIERREGENERAL} (apart from some minor additional terms), and thus is not
reproduced here. We are led to the identification of the zeroth order in $%
\mathbf{p,r}$
\begin{eqnarray}
\varepsilon _{\alpha +d\alpha }\left( \mathbf{P,R}\right) &=&\varepsilon
_{\alpha }\left( \mathbf{P,R}\right) +\left[ \partial _{\alpha }U_{\alpha
}\left( \mathbf{P},\mathbf{R}\right) U_{\alpha }^{+}\left( \mathbf{P,R}%
\right) ,\varepsilon _{\alpha }\left( \mathbf{P,R}\right) \right] d\alpha
\nonumber \\
&&+\frac{1}{2}\left\{ \mathcal{A}_{\alpha }^{R_{l}}\nabla
_{R_{l}}\varepsilon _{\alpha }\left( \mathbf{P,R}\right) +\nabla
_{R_{l}}\varepsilon _{\alpha }\left( \mathbf{P,R}\right) \mathcal{A}_{\alpha
}^{R_{l}}+\mathcal{A}_{\alpha }^{P_{l}}\nabla _{P_{l}}\varepsilon _{\alpha
}\left( \mathbf{P,R}\right) +\nabla _{P_{l}}\varepsilon _{\alpha }\left(
\mathbf{P,R}\right) \mathcal{A}_{\alpha }^{P_{l}}\right\} d\alpha  \nonumber
\\
&&+\frac{i}{2}\left\{ \left[ \varepsilon _{\alpha }\left( \mathbf{P,R}%
\right) ,\mathcal{A}_{\alpha }^{R_{l}}\right] \mathcal{A}_{\alpha
}^{P_{l}}d\alpha -\left[ \varepsilon _{\alpha }\left( \mathbf{P,R}\right) ,%
\mathcal{A}_{\alpha }^{P_{l}}\right] \mathcal{A}_{\alpha }^{R_{l}}-\left[
Asym\mathcal{A}_{\alpha }^{P_{l},R_{l}},\varepsilon _{\alpha }\left( \mathbf{%
P,R}\right) \right] \right\} d\alpha  \nonumber \\
&&+\frac{id\alpha }{2}\left\{ Asym\left\{ \nabla _{P_{l}}\nabla
_{R_{l}}\varepsilon _{\alpha }\left( \mathbf{P,R}\right) \right\} -U_{\alpha
}Asym\left\{ \nabla _{P_{l}}\nabla _{R_{l}}H_{0}\left( \mathbf{P,R}\right)
\right\} U_{\alpha }^{+}\right\}  \label{equa}
\end{eqnarray}
where we have defined the Berry phases \cite{PIERREGENERAL} (note the
absence of $\alpha $ in factor compared to the usual definition)
\begin{eqnarray*}
\mathcal{A}_{\alpha }^{R_{l}} &=&iU_{\alpha }\left( \mathbf{P},\mathbf{R}%
\right) \nabla _{P_{l}}U_{\alpha }^{+}\left( \mathbf{P},\mathbf{R}\right) \\
\mathcal{A}_{\alpha }^{P_{l}} &=&-iU_{\alpha }\left( \mathbf{P},\mathbf{R}%
\right) \nabla _{R_{l}}U_{\alpha }^{+}\left( \mathbf{P},\mathbf{R}\right)
\end{eqnarray*}
and the ''second order'' Berry phase
\[
\mathcal{A}_{\alpha }^{P_{l},R_{l}}=\left[ \nabla _{R_{k}}\nabla
_{P_{l}}U_{\alpha }\left( \mathbf{P},\mathbf{R}\right) \right] U_{\alpha
}^{+}\left( \mathbf{P},\mathbf{R}\right)
\]
We have moreover introduced the linear operation $Asym$ \cite{PIERREGENERAL}%
. It acts on a symmetrical function in $\mathbf{P}$ and $\mathbf{R}$ in the
following way :
\begin{equation}
Asym\left\{ \frac{1}{2}A\left( \mathbf{R}\right) B\left( \mathbf{P}\right) +%
\frac{1}{2}B\left( \mathbf{P}\right) A\left( \mathbf{R}\right) \right\} =%
\left[ B\left( \mathbf{P}\right) ,A\left( \mathbf{R}\right) \right]
\end{equation}
the functions $A\left( \mathbf{R}\right) $ and $B\left( \mathbf{P}\right) $
beeing typically monomials in $\mathbf{R}$ and $\mathbf{P}$ arising in the
series expansions of the physical quantities.

Note that the $Asym$ operator especially in the term $U_{\alpha
}^{+}Asym\left\{ \nabla _{P_{l}}\nabla _{R_{l}}H_{0}\left( \mathbf{P,R}%
\right) \right\} U_{\alpha }^{+}$, is not particularly elegant. It could in
fact be rewritten as an infinite series of differential operators, but the
gain does not seem to be clear at this point. Moreover, in our examples, our
choice will appear to be the most convenient.

Ultimately, we arrive at the following differential equation
\begin{eqnarray}
\frac{d}{d\alpha }\varepsilon _{\alpha }\left( \mathbf{P,R}\right) &=&\left[
\partial _{\alpha }U_{\alpha }\left( \mathbf{P},\mathbf{R}\right) U_{\alpha
}^{+}\left( \mathbf{P,R}\right) ,\varepsilon _{\alpha }\left( \mathbf{P,R}%
\right) \right]  \nonumber \\
&&+\left\{ \frac{1}{2}\mathcal{A}_{\alpha }^{R_{l}}\nabla
_{R_{l}}\varepsilon _{\alpha }\left( \mathbf{P,R}\right) +\nabla
_{R_{l}}\varepsilon _{\alpha }\left( \mathbf{P,R}\right) \mathcal{A}_{\alpha
}^{R_{l}}+\mathcal{A}_{\alpha }^{P_{l}}\nabla _{P_{l}}\varepsilon _{\alpha
}\left( \mathbf{P,R}\right) +\nabla _{P_{l}}\varepsilon _{\alpha }\left(
\mathbf{P,R}\right) \mathcal{A}_{\alpha }^{P_{l}}\right\}  \nonumber \\
&&+\frac{i}{2}\left\{ \left[ \varepsilon _{\alpha }\left( \mathbf{P,R}%
\right) ,\mathcal{A}_{\alpha }^{R_{l}}\right] \mathcal{A}_{\alpha }^{P_{l}}-%
\left[ \varepsilon _{\alpha }\left( \mathbf{P,R}\right) ,\mathcal{A}_{\alpha
}^{P_{l}}\right] \mathcal{A}_{\alpha }^{R_{l}}-\left[ Asym\mathcal{A}%
_{\alpha }^{P_{l},R_{l}},\varepsilon _{\alpha }\left( \mathbf{P,R}\right) %
\right] \right\}  \nonumber \\
&&+\frac{i}{2}\left\{ Asym\left\{ \nabla _{P_{l}}\nabla _{R_{l}}\varepsilon
_{\alpha }\left( \mathbf{P,R}\right) \right\} -U_{\alpha }Asym\left\{ \nabla
_{P_{l}}\nabla _{R_{l}}H_{0}\left( \mathbf{P,R}\right) \right\} U_{\alpha
}^{+}\right\}  \label{equalast}
\end{eqnarray}
To complete the diagonalization process, we have to couple this equation
with the evolution as a function of $\alpha $ of the transformation matrix $%
U_{\alpha }\left( \mathbf{P},\mathbf{R}\right) $. To find this last one, let
us combine Eqs. \ref{unmoinsUU} and \ref{UUplus} to get
\[
0=U_{\alpha }(\mathbf{P},\mathbf{R)}U_{\alpha }^{+}(\mathbf{P,R)-}U\mathbf{%
_{\alpha }(\mathbf{P+p},\mathbf{R+r)}}U\mathbf{_{\alpha }^{+}(\mathbf{%
P+p,R+r)}+}d\alpha \left[ \partial _{\alpha }U_{\alpha }(\mathbf{P},\mathbf{%
R)}U_{\alpha }^{+}(\mathbf{P,R)+}U\mathbf{_{\alpha }(\mathbf{P},\mathbf{R)}}U%
\mathbf{_{\alpha }^{+}(\mathbf{P,R)}}\right]
\]
This latter expression can be, similarly to the energy, expanded as
\begin{equation}
0=\partial _{\alpha }U_{\alpha }(\mathbf{P},\mathbf{R)}U_{\alpha }^{+}(%
\mathbf{P,R)+}U\mathbf{_{\alpha }(\mathbf{P},\mathbf{R)}}\partial _{\alpha }U%
\mathbf{_{\alpha }^{+}(\mathbf{P,R)}}-\frac{i}{2}Asym\left( \mathcal{A}%
_{\alpha }^{P_{l},R_{l}}+\left( \mathcal{A}_{\alpha }^{P_{l},R_{l}}\right)
^{+}\right) -\frac{i}{2}\left[ \mathcal{A}_{\alpha }^{R_{l}},\mathcal{A}%
_{\alpha }^{P_{l}}\right]  \label{Uevolution}
\end{equation}
With these two equations Eqs. \ref{equalast} and \ref{Uevolution} at hand,
our diagonalization process can be performed. Actually, since all quantities
are matrix valued and since $\varepsilon _{\alpha }\left( \mathbf{P,R}%
\right) $ is by definition a diagonal matrix, we can separate the energy
equation Eq. \ref{equalast} in a diagonal and a non diagonal part such that
we are led to the following two equations
\begin{eqnarray}
\frac{d}{d\alpha }\varepsilon _{\alpha }\left( \mathbf{P,R}\right) &=&%
\mathcal{P}_{+}[\text{R.H.S}.\text{ of Eq. \ref{equalast}}]  \label{eq1} \\
0 &=&\mathcal{P}_{-}[\text{R.H.S}.\text{ of Eq. \ref{equalast}}]  \label{eq2}
\end{eqnarray}
where $\mathcal{P}_{+}[...]$ and $\mathcal{P}_{-}[...]$ denote the
projection on the diagonal and non-diagonal part respectively. These two
equations are supplemented by the differential unitarity condition
\begin{equation}
0=\partial _{\alpha }U_{\alpha }(\mathbf{P},\mathbf{R)}U_{\alpha }^{+}(%
\mathbf{P,R)+}U\mathbf{_{\alpha }(\mathbf{P},\mathbf{R)}}\partial _{\alpha }U%
\mathbf{_{\alpha }^{+}(\mathbf{P,R)}}-\frac{i}{2}Asym\left( \mathcal{A}%
_{\alpha }^{P_{l},R_{l}}+(\mathcal{A}_{\alpha }^{P_{l},R_{l}})^{+}\right) -%
\frac{i}{2}\left[ \mathcal{A}_{\alpha }^{R_{l}},\mathcal{A}_{\alpha }^{P_{l}}%
\right]  \label{eq3}
\end{equation}
We claim that those three equations Eqs.\ref{eq1}-\ref{eq3} allow to
determine recursively in powers of $\alpha $ the energy of the quantum
system in question. Actually, the integration over $\alpha $ of Eq. \ref{eq1}
gives $\varepsilon _{\alpha }\left( \mathbf{P,R}\right) $ at order $n$ in $%
\alpha $ when knowing all quantities at order $n-1$. By the same token, Eqs.
\ref{eq2} and \ref{eq3} (whose meaning is that $U_{\alpha }\left( \mathbf{P},%
\mathbf{R}\right) $ is unitary at each order in $\alpha $) involve $\partial
_{\alpha }U_{\alpha }\left( \mathbf{P},\mathbf{R}\right) $, and allow to
recover $U_{\alpha }\left( \mathbf{P},\mathbf{R}\right) $ at order $n$ by
integration over $\alpha $. Note however that these two equations allow only
fixing\ partially $U_{\alpha }\left( \mathbf{P},\mathbf{R}\right) $, leaving
$m$ real parameters free with $m$ the size of the matrix. This reflects a
kind of gauge invariance (we will see a practical example of this in section
four). Actually, the diagonalization matrix is not unique, so that we are
left with some choice of the parameters. For example one can impose the
diagonal elements to be real (see below in the example). As a consequence,
the diagonalization process is perfectly controlled order by order in the
series expansion in $\alpha $.

\section{The semiclassical approximation}

In this section we consider the Hamiltonian diagonalization at the
semiclassical level and the resulting equations of motion. Actually, the
semiclassical approximation has recently found new important applications in
particle and solid state physics. Notably, the equations of motion reveal a
new contribution coming from the Berry curvature. This contribution, called
the anomalous velocity, modifies profoundly the dynamics of the particles.
For instance, the spin Hall effect of electrons and holes in semiconductors
\cite{MURAKAMI1}, as well as the new discovered optical Hall effect \cite
{MURAKAMI2}\cite{ALAIN}\cite{BLIOKH2}\cite{PIERREPHOTON} can be interpreted
in this context. Similarly, the recent experimental discovery of the
monopole in momentum can also be elegantly interpreted as the influence of
the Berry curvature on the semiclassical dynamics of Bloch electrons \cite
{FANG}\cite{ALAINMONOPOLE}.

\subsection{The semiclassical energy}

The consideration of Eq. \ref{equalast} alone is sufficient to deduce the
semiclassical diagonal Hamiltonian. Indeed, writing $\varepsilon _{\alpha
}=\varepsilon _{0}+\alpha \varepsilon _{1},$ with $\varepsilon _{0}$ the
diagonalized energy at the zero order, Eq. \ref{equalast} is solved by
(putting $\alpha =\hbar $)
\begin{eqnarray}
\varepsilon \left( \mathbf{P,R}\right) &=&\varepsilon _{0}\left( \mathbf{P,R}%
\right) +\hbar \left\{ \frac{1}{2}\emph{A}_{0}^{R_{l}}\nabla
_{R_{l}}\varepsilon _{0}\left( \mathbf{P,R}\right) +\nabla
_{R_{l}}\varepsilon _{0}\left( \mathbf{P,R}\right) \emph{A}_{0}^{R_{l}}+%
\emph{A}_{0}^{P_{l}}\nabla _{P_{l}}\varepsilon _{0}\left( \mathbf{P,R}%
\right) \right.  \nonumber \\
&&+\left. \nabla _{P_{l}}\varepsilon _{0}\left( \mathbf{P,R}\right) \emph{A}%
_{0}^{P_{l}}\right\} +\frac{i\hbar }{2}\mathcal{P}_{+}\left\{ \left[
\varepsilon _{0}\left( \mathbf{P,R}\right) ,\mathcal{A}_{0}^{R_{l}}\right]
\mathcal{A}_{0}^{P_{l}}-\left[ \varepsilon _{0}\left( \mathbf{P,R}\right) ,%
\mathcal{A}_{0}^{P_{l}}\right] \mathcal{A}_{0}^{R_{l}}\right\}
\label{semiclassicalenergy}
\end{eqnarray}
where we have introduced the notations $\emph{A}_{0}^{\mathbf{R}}=\mathcal{P}%
_{+}\left[ \mathcal{A}_{0}^{\mathbf{R}}\right] $ and $\emph{A}_{0}^{\mathbf{P%
}}=\mathcal{P}_{+}\left[ \mathcal{A}_{0}^{\mathbf{P}}\right] $.

This latter expression can also be written
\begin{equation}
\varepsilon \left( \mathbf{P,R}\right) \simeq \varepsilon \left( \mathbf{p,r}%
\right) +\frac{i\hbar }{2}\mathcal{P}_{+}\left[ \left[ \varepsilon \left(
\mathbf{p,r}\right) ,\mathcal{A}_{0}^{R_{l}}\right] \mathcal{A}_{0}^{P_{l}}-%
\left[ \varepsilon \left( \mathbf{p,r}\right) ,\mathcal{A}_{0}^{P_{l}}\right]
\mathcal{A}_{0}^{R_{l}}\right] +O(\hbar ^{2})  \label{eqenersc}
\end{equation}
where we have defined the projected dynamical operators (covariant
coordinates and momentum operators)
\begin{eqnarray}
\mathbf{r} &=&\mathbf{R}+\hbar \emph{A}_{0}^{\mathbf{R}}  \nonumber \\
\mathbf{p} &=&\mathbf{P}+\hbar \emph{A}_{0}^{\mathbf{P}}
\end{eqnarray}
with $\mathcal{A}_{0}^{\mathbf{R}}=i\left[ U_{0}\nabla _{\mathbf{P}}U_{0}^{+}%
\right] $, $\mathcal{A}_{0}^{\mathbf{P}}=-i\left[ U_{0}\nabla _{\mathbf{R}%
}U_{0}^{+}\right] ,$and $\mathcal{A}_{0}^{\mathbf{P,R}}=\left[ \nabla _{%
\mathbf{R}}\nabla _{\mathbf{P}}U_{0}\right] U_{0}^{+}$.

The matrix $U_{0}\left( \mathbf{P,R}\right) $ is the diagonalization matrix
for $H_{0}$ when the operators are supposed to be commuting quantities, the
diagonalized energy being $\varepsilon _{0}\left( \mathbf{P,R}\right) .$
When $\mathbf{P}$ and $\mathbf{R}$ do not commute, the matrix $U_{0}\left(
\mathbf{P,R}\right) $ does not diagonalize $H_{0}$ anymore. In order to get
the corrections to the energy at the semiclassical order due to the
noncommutativity of $\mathbf{P}$ and $\mathbf{R}$ we have to compute $%
\varepsilon _{1}\left( \mathbf{P,R}\right) $. Note that in Eq. \ref
{semiclassicalenergy}, the symmetrization defined above (that is all the
powers of $\mathbf{P}$ have been put half on the left and half of the right
in each expression) is assumed. From the diagonal Hamiltonian, we can now
derive the equations of motion for the covariant operators.

\subsection{ The equations of motion}

Given the Hamiltonian derived in the previous subsection, the equations of
motion can now be easily derived. As usual, the dynamics equations have to
be considered, not for the usual position and momentum, but rather for the
projected variables $\mathbf{r}$ and $\mathbf{p}$ . Actually, these latter
naturally appear in our diagonalization process at the $\hbar $ order. Let
us remark, as now well known, that their components do not commute any more.
Actually (removing the index $\alpha $ for clarity)
\begin{eqnarray}
\left[ r_{i},r_{j}\right] &=&i\hbar ^{2}\Theta _{ij}^{rr}=i\hbar ^{2}\left(
\nabla _{P_{i}}\emph{A}_{R_{j}}-\nabla _{P_{j}}\emph{A}_{R_{i}}\right)
+\hbar ^{2}\left[ \emph{A}_{R_{j}},\emph{A}_{R_{i}}\right]  \nonumber \\
\left[ p_{i},p_{j}\right] &=&i\hbar ^{2}\Theta _{ij}^{pp}=-i\hbar ^{2}\left(
\nabla _{R_{i}}\emph{A}_{P_{j}}-\nabla _{R_{j}}\emph{A}_{P_{i}}\right)
+\hbar ^{2}\left[ \emph{A}_{P_{i}},\emph{A}_{P_{j}}\right]  \nonumber \\
\left[ p_{i},r_{j}\right] &=&-i\hbar \delta _{ij}+i\hbar ^{2}\Theta
_{ij}^{pr}=-i\hbar \delta _{ij}-i\hbar ^{2}\left( \nabla _{R_{i}}\emph{A}%
_{R_{j}}+\nabla _{P_{j}}\emph{A}_{P_{i}}\right) +\hbar ^{2}\left[ \emph{A}%
_{P_{i}},\emph{A}_{R_{j}}\right]
\end{eqnarray}
the $\Theta _{ij}$ being the so called Berry curvatures.

Using now our Hamiltonian yields directly to general equations of motion for
$\mathbf{r}$, $\mathbf{p}$ :
\begin{eqnarray}
\mathbf{\dot{r}} &=&\frac{i}{\hbar }\left[ \mathbf{r},\varepsilon \left(
\mathbf{p,r}\right) \right] +\frac{i}{\hbar }\left[ \mathbf{r},\frac{i\hbar
}{2}\mathcal{P}_{+}\left[ \left[ \varepsilon \left( \mathbf{p,r}\right) ,%
\emph{A}_{R_{l}}\right] \emph{A}_{P_{l}}-\left[ \varepsilon \left( \mathbf{%
p,r}\right) ,\emph{A}_{P_{l}}\right] \emph{A}_{R_{l}}\right] \right]
\nonumber \\
\mathbf{\dot{p}} &=&\frac{i}{\hbar }\left[ \mathbf{p},\varepsilon \left(
\mathbf{p,r}\right) \right] +\frac{i}{\hbar }\left[ \mathbf{p},\frac{i\hbar
}{2}\mathcal{P}_{+}\left[ \left[ \varepsilon \left( \mathbf{p,r}\right) ,%
\emph{A}_{R_{l}}\right] \emph{A}_{P_{l}}-\left[ \varepsilon \left( \mathbf{%
p,r}\right) ,\emph{A}_{P_{l}}\right] \emph{A}_{R_{l}}\right] \right]
\end{eqnarray}
The commutators can be computed through the previous commutation rules
between $\mathbf{r}$ and $\mathbf{p}$. The last term in each equation
represents a contribution of ''magnetization'' type and has the advantage to
present this general form whatever the system initially considered. In the
context of Bloch electrons in a magnetic field, it gives exactly the
magnetization term revealed in \cite{NIU1} (see \cite{PIERREBLOCH}). For
spinning particles in static gravitational fields, this term gives a
coupling between the spin and the intrinsic angular momentum with
magneto-torsion fields \cite{PIERREPHOTON}\cite{PIERREELECTRON}. It is
interesting to note that both the Hamiltonian Eq. \ref{eqenersc} and the
equations of motion have already been derived in article \cite{PIERREGENERAL}
.Our method appears then to be an extension of \cite{PIERREGENERAL}.
However, although relying on this previous method, it seems to us more
elegant, and offers to possibility to go beyond the semiclassical level.
This will appear very important in the next section to compute the velocity
of a spinning massless particle (photon) in an inhomogeneous medium. The
second application is the determination of the equations of motion of a
Bloch electron in an external constant electric field. In both situations we
are going to consider the second order in $\hbar $ quantum effects.

\section{Physical applications}

\subsection{Massless spinning particles in an isotropic inhomogeneous medium.%
}

We consider here the dynamics of a spinning massless particle propagating in
a isotropic inhomogeneous medium of index $n(\mathbf{R})$. We thus start
with the Hamiltonian of a massless Dirac particle in an curved space of
metric $g^{ij}(\mathbf{R})=n^{-1}\left( \mathbf{R}\right) \delta ^{ij}$ as
in \cite{HORVATHY1}\cite{PIERREPHOTON}. The case of higher spin (like the
photon) can be treated in the same manner through the Bargman-Wigner
equations as in \cite{PIERREPHOTON}. For this reason we limit ourself to the
Dirac particles (with the convention $c=1$) and write the Hamiltonian as
\begin{equation}
H_{0}=\frac{1}{2}\left( \mathbf{\alpha }.\mathbf{P}F(\mathbf{R})+F(\mathbf{R}%
)\mathbf{\alpha }.\mathbf{P}\right)
\end{equation}
with $F(\mathbf{R})=n^{-1}\left( \mathbf{R}\right) $. Let us start to
recover the first order diagonalization given in \cite{PIERREPHOTON}. To
exploit the previous general formula we write all quantities as a series
expansion in $\alpha ,$ thus $\varepsilon _{\alpha }=\varepsilon _{0}+\alpha
\varepsilon _{1}+...$, $U_{\alpha }=U_{0}+\alpha U_{1}+...,\mathcal{A}%
_{\alpha }^{R}=\mathcal{A}_{0}^{R}+\alpha \mathcal{A}_{1}^{R}+...$ and the
same for all expressions. In these expressions, all terms are functions of $%
\mathbf{P}$ and $\mathbf{R}$ and are thus assumed to be symmetrized in the
sense defined above.

\subsubsection{First order (semiclassical) diagonalization.}

We already know (see \cite{PIERREPHOTON}) that at the zeroth order in $%
\alpha $, that is when $\mathbf{R}$ and $\mathbf{P}$ commute, the
Hamiltonian diagonalization can be performed through the following
Foldy-Wouthuysen unitary matrix
\begin{equation}
U_{0}\left( \mathbf{P}\right) =\frac{\sqrt{\mathbf{P}^{2}}+\beta \widetilde{%
\mathbf{\alpha }}\mathbf{.P}}{\sqrt{2\mathbf{P}^{2}}}  \label{U0}
\end{equation}
(where $\beta $ and $\widetilde{\mathbf{\alpha }}$ are the usual notations
for the ''Dirac matrices'' ) such that in the diagonal representation the
Hamiltonian becomes
\begin{equation}
U_{0}H_{0}\left( \mathbf{P,R}\right) U_{0}^{+}=\varepsilon _{0}\left(
\mathbf{P,R}\right) =\frac{1}{2}\left( \beta F(\mathbf{R)}\sqrt{\mathbf{P}%
^{2}}+\beta \sqrt{\mathbf{P}^{2}}F(\mathbf{R)}\right)
\end{equation}
From the unitary matrix, we can deduce that at this order, only $\mathcal{A}%
_{0}^{\mathbf{R}}=i\left[ U\nabla _{\Bbb{P}}U^{+}\right] $ is non zero, thus
$\mathcal{A}_{0}^{\mathbf{P}}=-i\hbar \left[ U\nabla _{\mathbf{R}}U^{+}%
\right] =0+O\left( \alpha \right) ,$and $\mathcal{A}_{0}^{P_{l},R_{l}}=0+O%
\left( \alpha \right) .$ Obviously, when $\mathbf{P}$ and $\mathbf{R}$ do
not commute, the matrix $\varepsilon _{0}\left( \mathbf{P,R}\right) $ does
not diagonalize $H_{0}$ anymore and one has to go to higher order in the
expansion in $\alpha .$

Let thus consider the diagonalization at order $\alpha .$ The differential
equation Eq. \ref{equalast} at the zeroth order (at this order, we consider
only in the right hand side of this equation the terms of order $\alpha ^{0}$%
) reduces to
\begin{equation}
\frac{d}{d\alpha }\varepsilon _{\alpha }\left( \mathbf{P,R}\right) =\frac{1}{%
2}\left[ \mathcal{A}_{0}^{R_{l}}\nabla _{R_{l}}\varepsilon _{0}\left(
\mathbf{P,R}\right) +\nabla _{R_{l}}\varepsilon _{0}\left( \mathbf{P,R}%
\right) \mathcal{A}_{0}^{R_{l}}\right] +\left[ U_{1}\left( \mathbf{P},%
\mathbf{R}\right) U_{0}^{+}\left( \mathbf{P}\right) ,\varepsilon _{0}\left(
\mathbf{P,R}\right) \right]
\end{equation}
and can be divided in two parts
\begin{equation}
\frac{d}{d\alpha }\varepsilon _{\alpha }\left( \mathbf{P,R}\right) =\frac{1}{%
2}\mathcal{P}_{+}\left[ \mathcal{A}_{0}^{R_{l}}\nabla _{R_{l}}\varepsilon
_{0}\left( \mathbf{P,R}\right) +\nabla _{R_{l}}\varepsilon _{0}\left(
\mathbf{P,R}\right) \mathcal{A}_{0}^{R_{l}}\right]  \label{eq1photon}
\end{equation}
and
\begin{equation}
0=\frac{1}{2}\mathcal{P}_{-}\left[ \mathcal{A}_{0}^{R_{l}}\nabla
_{R_{l}}\varepsilon _{0}\left( \mathbf{P,R}\right) +\nabla
_{R_{l}}\varepsilon _{0}\left( \mathbf{P,R}\right) \mathcal{A}_{0}^{R_{l}}%
\right] +\left[ U_{1}\left( \mathbf{P},\mathbf{R}\right) U_{0}^{+}\left(
\mathbf{P}\right) ,\varepsilon _{0}\left( \mathbf{P,R}\right) \right]
\label{eq2photon}
\end{equation}
(the last term being always anti-diagonal at this order). By integrating Eq.
\ref{eq1photon} with respect to $\alpha $ from $o$ to $\hbar $ the first
equation, we get the energy operator at the first order in $\alpha =\hbar $
which reads
\[
\varepsilon _{\alpha }\left( \mathbf{P,R}\right) =\varepsilon _{0}\left(
\mathbf{P,R}\right) +\frac{\hbar }{2}\emph{A}_{0}^{R_{l}}\nabla
_{R_{l}}\varepsilon _{0}\left( \mathbf{P,R}\right) +\frac{\hbar }{2}\nabla
_{R_{l}}\varepsilon _{0}\left( \mathbf{P,R}\right) \emph{A}_{0}^{R_{l}}
\]
The energy $\varepsilon _{\alpha }$ can be recombined under the usual form
\cite{PIERREGENERAL}\cite{PIERREPHOTON} and written in term of the covariant
coordinate operator $\mathbf{r}^{+}$ as
\begin{equation}
\varepsilon _{\alpha }\left( \mathbf{P,r}\right) =\frac{\beta }{2}\left(
F\left( \mathbf{r}\right) \sqrt{\mathbf{P}^{2}}+\sqrt{\mathbf{P}^{2}}F\left(
\mathbf{r}\right) \right)
\end{equation}
with
\begin{equation}
\mathbf{r}=\mathbf{R+}\hbar \emph{A}_{0}^{\mathbf{R}}=\mathbf{R+}i\hbar
\frac{\mathbf{P}\times \mathbf{\Sigma }}{2\mathbf{P}^{2}}
\end{equation}
where $\emph{A}_{0}^{\mathbf{R}}$ means that we need only the diagonal part
of $\mathcal{A}_{0}^{\mathbf{R}}.$ The matrix $\mathbf{\Sigma =1\otimes
\sigma }$ is a $4\times 4$ matrix.

From the last two equations we can deduce the equations of motion in the
semiclassical approximation. Indeed, for a particle of positive energy only $%
\varepsilon \left( \mathbf{P,r}\right) =\frac{1}{2}\left( F\left( \mathbf{r}%
\right) \sqrt{\mathbf{P}^{2}}+\sqrt{\mathbf{P}^{2}}F\left( \mathbf{r}\right)
\right) $ with now $\mathbf{r}=\mathbf{R+}i\frac{\mathbf{P}\times \mathbf{S}%
}{\mathbf{P}^{2}}$ a $2\times 2$ matrix (the spin matrix is $\mathbf{S=\hbar
\sigma /2}$), the usual relations $\mathbf{r}=-\frac{i}{\hbar }\left[
\mathbf{r},\varepsilon \left( \mathbf{P,r}\right) \right] $ and $\mathbf{P}=-%
\frac{i}{\hbar }\left[ \mathbf{P},\varepsilon \left( \mathbf{P,r}\right) %
\right] $, lead to the by now well known equations of motion
\begin{eqnarray}
\mathbf{\dot{r}} &=&\nabla _{\mathbf{P}}\varepsilon +\hbar \mathbf{\dot{P}%
\times }\Theta ^{rr}  \nonumber \\
\mathbf{\dot{P}} &=&\nabla _{\mathbf{r}}\varepsilon  \label{eqmotion}
\end{eqnarray}
where $\left[ r_{i},r_{j}\right] =i\hbar ^{2}\Theta _{ij}^{rr}=i\hbar
^{2}\varepsilon _{ijk}\Theta _{k}^{rr}=-i\hbar \varepsilon _{ijk}\lambda
\frac{P^{k}}{P^{3}}$, and $\lambda =\mathbf{S.P/}P$ the helicity. These
latter equations are very important because they induce the optical Hall
effect or spin Hall effect of light (for the photon one just has to replace
the Pauli matrices $\mathbf{\sigma }$ by the spin one matrices) as a
consequence of the anomalous velocity term $\hbar \mathbf{\dot{P}\times }%
\Theta ^{rr}$. This effect has been discussed in several circumstances \cite
{MURAKAMI2}\cite{BLIOKH2} \cite{HORVATHY1} and was considered in a quantum
mechanical context in \cite{PIERREGENERAL}\cite{ALAIN}\cite{PIERREPHOTON}.
From Eq.\ref{eqmotion} we see that the velocity components are given by the
relation
\begin{equation}
v^{i}=\frac{1}{2}\left( \frac{c}{n(\mathbf{r})}\frac{P^{i}}{P}+\frac{P^{i}}{P%
}\frac{c}{n(\mathbf{r})}\right) +\frac{\lambda }{P^{2}}\frac{c}{n(\mathbf{r})%
}\varepsilon _{ijk}P^{k}\frac{\partial \ln n}{\partial x^{i}}
\label{velocity}
\end{equation}
Note that the symmetrization for the last term was unnecessary due to the
presence of the $\hbar $. Symmetrizing would just add $\hbar ^{2}$
corrections.

From the last relation, we deduce the velocity
\begin{equation}
v=\frac{c}{n(\mathbf{r})}\left( 1+\frac{\lambda ^{2}}{P^{2}}\left( \left(
\nabla \ln n\right) ^{2}-\frac{1}{P^{2}}\left( \mathbf{P.}\nabla \ln
n\right) ^{2}\right) \right) ^{1/2}  \label{velo}
\end{equation}
The velocity can be rewritten $v=\frac{c}{n(\mathbf{r})}\left( 1+\frac{%
\lambda ^{2}}{P^{2}}\left( \nabla \ln n\right) ^{2}\sin ^{2}\vartheta
\right) ^{1/2}$ with $\vartheta $ the angle between $\mathbf{P}$ and $\nabla
\ln n$. We therefore find the important result that the light velocity in an
isotropic inhomogeneous medium does not have the well known expression $v(%
\mathbf{r})=c/n(\mathbf{r})$ but has rather quantum corrections due to the
spin-orbit couplings when $\mathbf{P}$and $\nabla \ln n$ are not parallel.
However, at this level of approximation, we can not yet conclude regarding
the velocity since the quantum corrections are of order $\hbar ^{2}$ which
is beyond the first order approximation. To get a clear answer one has to
compute the Hamiltonian diagonalization at the second order in $\hbar $.

\subsubsection{Second order diagonalization}

To perform the diagonalization at order $\alpha ^{2}$ we need to know the
transformation matrix $U_{\alpha }$ at the first order in $\alpha $, i.e.,
the matrix $U_{1}(\mathbf{P},\mathbf{R)}$ (i.e. the Foldy Fouthuysen
transformation when $\mathbf{R}$ and $\mathbf{P}$ do not commute). The
equation in Eq. \ref{eq2photon} allows to find easily the transformation
matrix $U_{\alpha }$ at this order. Indeed, given that at zeroth order
\begin{equation}
\mathcal{P}_{-}\left[ \mathcal{A}_{0}^{R_{l}}\nabla _{R_{l}}\varepsilon
_{0}\left( \mathbf{P,R}\right) +\nabla _{R_{l}}\varepsilon _{0}\left(
\mathbf{P,R}\right) \mathcal{A}_{0}^{R_{l}}\right] =0
\end{equation}
as a consequence of $\widetilde{\mathbf{\alpha }}\beta +\beta \widetilde{%
\mathbf{\alpha }}=0$, we can write
\begin{equation}
\left[ U_{1}\left( \mathbf{P},\mathbf{R}\right) U_{0}^{+}\left( \mathbf{P}%
\right) ,\varepsilon _{0}\left( \mathbf{P,R}\right) \right] =0  \label{U1}
\end{equation}
Likewise, the differential unitarity condition Eq. \ref{eq3} at the same
zero order becomes
\begin{equation}
U_{1}(\mathbf{P},\mathbf{R)}U_{0}^{+}(\mathbf{P)+}U\mathbf{_{0}(\mathbf{P)}}U%
\mathbf{_{1}^{+}(\mathbf{P,R)=}}0  \label{U2}
\end{equation}
We claim that $U_{\alpha }\left( \mathbf{P},\mathbf{R}\right) $, and
consequently the Berry phases, are unchanged with respect to the zeroth
order. That is the matrix Eq. \ref{U0} also diagonalizes $H_{0}$ at the
first order in $\alpha $. Indeed, we can easily see with the help of Eqs.
\ref{U1}\ref{U2}, that the following transformation $\left( 1-\alpha U_{1}(%
\mathbf{P},\mathbf{R)}U_{0}^{+}(\mathbf{P)}\right) U_{\alpha }(\mathbf{P},%
\mathbf{R)}$ which at order $\alpha $ is equal to $U_{0}(\mathbf{P)}$
diagonalizes $H_{0}$, e.g. $U_{0}^{+}H_{0}U_{0}=\varepsilon _{0}\left(
\mathbf{P,R}\right) +\alpha \varepsilon _{1}\left( \mathbf{P,R}\right) $
(this is an example of the gauge transformation in the diagonalization
matrix we mentioned previously).

As a consequence, the Berry phases do not get any contribution at the first
order. Therefore we have $\emph{A}_{\alpha }^{\mathbf{P}}=-i\mathcal{P}_{+}%
\left[ U\nabla _{\mathbf{R}}U^{+}\right] =0+O\left( \alpha ^{2}\right) $
implying $\emph{A}_{\alpha }^{P_{l},R_{l}}=0+O\left( \alpha ^{2}\right) $
and also
\begin{equation}
\emph{A}_{\alpha }^{\mathbf{R}}=i\hbar \mathcal{P}_{+}\left[ U\nabla
_{P}U^{+}\right] =i\frac{\mathbf{P}\times \mathbf{\Sigma }}{2\mathbf{P}^{2}}%
+O\left( \alpha ^{2}\right)
\end{equation}
Now, to find the second order Hamiltonian we write the differential equation
Eq. \ref{equalast} at the first order.
\begin{equation}
\frac{d}{d\alpha }\varepsilon _{\alpha }\left( \mathbf{P,R}\right) =\frac{1}{%
2}\mathcal{P}_{+}\left[ \mathcal{A}_{\alpha }^{R_{l}}\nabla
_{R_{l}}\varepsilon _{\alpha }\left( \mathbf{P,R}\right) +\nabla
_{R_{l}}\varepsilon _{\alpha }\left( \mathbf{P,R}\right) \mathcal{A}_{\alpha
}^{R_{l}}\right] +\frac{i}{2}Asym\left\{ \nabla _{P_{l}}\nabla
_{R_{l}}\varepsilon _{\alpha }\left( \mathbf{P,R}\right) \right\}
\label{eq222}
\end{equation}
since trivially $Asym\left\{ \nabla _{P_{l}}\nabla _{R_{l}}H_{0}\left(
\mathbf{P,R}\right) \right\} =0.$ To show how to solve this equation, recall
that at the first order $\varepsilon _{\alpha }\left( \mathbf{P,R}\right) $
is given by
\begin{equation}
\varepsilon _{\alpha }\left( \mathbf{P,R}\right) =\varepsilon _{0}\left(
\mathbf{P,R}\right) +\frac{\alpha }{2}\emph{A}_{\alpha }^{R_{l}}\nabla
_{R_{l}}\varepsilon _{\alpha }\left( \mathbf{P,R}\right) +\frac{\alpha }{2}%
\nabla _{R_{l}}\varepsilon _{\alpha }\left( \mathbf{P,R}\right) \emph{A}%
_{\alpha }^{R_{l}}
\end{equation}
(recall that all our expressions are supposed symmetrized in $\mathbf{P}$
and $\mathbf{R}$ as explained in the beginning of the previous section), so
that Eq. \ref{eq222} becomes
\begin{eqnarray*}
\frac{d}{d\alpha }\varepsilon _{\alpha }\left( \mathbf{P,R}\right) &=&\frac{1%
}{2}\emph{A}_{\alpha }^{R_{l}}\nabla _{R_{l}}\varepsilon _{0}\left( \mathbf{%
P,R}\right) +\frac{1}{2}\nabla _{R_{l}}\varepsilon _{0}\left( \mathbf{P,R}%
\right) \emph{A}_{\alpha }^{R_{l}}+\frac{\alpha }{4}\emph{A}_{\alpha
}^{R_{k}}\emph{A}_{\alpha }^{R_{l}}\nabla _{R_{k}}\nabla _{R_{l}}\varepsilon
_{0}\left( \mathbf{P,R}\right) \\
&&+\frac{\alpha }{2}\emph{A}_{\alpha }^{R_{k}}\nabla _{R_{k}}\nabla
_{R_{l}}\varepsilon _{0}\left( \mathbf{P,R}\right) \emph{A}_{\alpha
}^{R_{l}}+\frac{\alpha }{4}\nabla _{R_{k}}\nabla _{R_{l}}\varepsilon
_{0}\left( \mathbf{P,R}\right) \emph{A}_{\alpha }^{R_{k}}\emph{A}_{\alpha
}^{R_{l}}+\frac{i}{2}Asym\left\{ \nabla _{P_{l}}\nabla _{R_{l}}\varepsilon
_{0}\left( \mathbf{P,R}\right) \right\}
\end{eqnarray*}
and as a result after integration,
\begin{eqnarray*}
\varepsilon _{\alpha }\left( \mathbf{P,R}\right) -\varepsilon _{0}\left(
\mathbf{P,R}\right) &=&\frac{\alpha }{2}\emph{A}_{\alpha }^{R_{l}}\nabla
_{R_{l}}\varepsilon _{0}\left( \mathbf{P,R}\right) +\frac{\alpha }{2}\nabla
_{R_{l}}\varepsilon _{0}\left( \mathbf{P,R}\right) \emph{A}_{\alpha
}^{R_{l}}+\frac{\alpha ^{2}}{8}\emph{A}_{\alpha }^{R_{k}}\emph{A}_{\alpha
}^{R_{l}}\nabla _{R_{k}}\nabla _{R_{l}}\varepsilon _{0}\left( \mathbf{P,R}%
\right) \\
&&+\frac{\alpha ^{2}}{4}\emph{A}_{\alpha }^{R_{k}}\nabla _{R_{k}}\nabla
_{R_{l}}\varepsilon _{0}\left( \mathbf{P,R}\right) \emph{A}_{\alpha
}^{R_{l}}+\frac{\alpha ^{2}}{8}\nabla _{R_{k}}\nabla _{R_{l}}\varepsilon
_{0}\left( \mathbf{P,R}\right) \emph{A}_{\alpha }^{R_{k}}\emph{A}_{\alpha
}^{R_{l}} \\
&&+\frac{i}{2}\int_{0}^{\alpha }d\alpha Asym\left\{ \nabla _{P_{l}}\nabla
_{R_{l}}\varepsilon _{0}\left( \mathbf{P,R}\right) \right\}
\end{eqnarray*}
Given our symmetrization convention, this last expression $\varepsilon
_{\alpha }\left( \mathbf{P,R}\right) $ can be shown to be equal to $%
\varepsilon _{0}\left( \mathbf{P,r}\right) $.

Given this last result, we can solve our differential equation at the second
order approximation. Actually, recalling that $\varepsilon _{0}\left(
\mathbf{P,R}\right) =\frac{1}{2}\left( \beta F(\mathbf{R)}\sqrt{\mathbf{P}%
^{2}}+\beta \sqrt{\mathbf{P}^{2}}F(\mathbf{R)}\right) $, one can first state
that :
\[
Asym\left\{ \nabla _{P_{l}}\nabla _{R_{l}}\varepsilon _{0}\left( \mathbf{P,R}%
\right) \right\} =\left[ \sqrt{\mathbf{P}^{2}},F\left( \mathbf{r}\right) %
\right]
\]
so that :
\[
\frac{i}{2}\int_{0}^{\alpha }d\alpha Asym\left\{ \nabla _{P_{l}}\nabla
_{R_{l}}\varepsilon _{0}\left( \mathbf{P,R}\right) \right\} =-\frac{1}{4}%
\hbar \left[ \sqrt{\mathbf{P}^{2}},F\left( \mathbf{r}\right) \right]
\]

Now, by replacing $\alpha $ by $\hbar $ and projecting on the positive
energy subspace, our solution is :.
\[
\varepsilon \left( \mathbf{P,r}\right) =\frac{1}{2}\left( F\left( \mathbf{r}%
\right) \sqrt{\mathbf{P}^{2}}+\sqrt{\mathbf{P}^{2}}F\left( \mathbf{r}\right)
\right) -\frac{1}{4}\hbar \left[ \sqrt{\mathbf{P}^{2}},F\left( \mathbf{r}%
\right) \right]
\]
which can be also written
\[
\varepsilon \left( \mathbf{P,r}\right) =\frac{1}{2}\left( F\left( \mathbf{r}%
\right) \sqrt{\mathbf{P}^{2}}+\sqrt{\mathbf{P}^{2}}F\left( \mathbf{r}\right)
\right) -\frac{1}{4}\frac{\hbar ^{2}}{\sqrt{\mathbf{P}^{2}}}\mathbf{P\nabla }%
F\left( \mathbf{r}\right)
\]
(the symmetrization of the last term is not necessary due to the presence of
the $\hbar ^{2}$.)

Now, concerning the equation of motion at the second order, their form is
only formally identical to Eq. \ref{eqmotion}. Actually, the energy has been
changed by getting quantum corrections of order $\hbar ^{2}$. Moreover, the
term $\hbar \mathbf{\dot{P}\times }\Theta ^{rr}$ in Eq.\ref{eqmotion} has to
be understood as beeing symetrized between $\mathbf{P}$ and $\mathbf{r}$.

Let us conclude this section by noting from the equations of motion that the
speed of light get also quantum correction of order $\hbar ^{2}$. Indeed,
one has now a correction to Eq. \ref{velocity} of order $\hbar ^{2}$
(remember that $\lambda $ is of order $\hbar $) such that the velocity
components become
\[
v^{i}=\frac{1}{2}\left( \frac{c}{n(\mathbf{r})}\frac{P^{i}}{P}+\frac{P^{i}}{P%
}\frac{c}{n(\mathbf{r})}\right) +\frac{\lambda }{2P^{2}}\varepsilon
_{ijk}\left( P^{k}\frac{\partial \ln n}{\partial x^{i}}\frac{c}{n(\mathbf{r})%
}+\frac{\partial \ln n}{\partial x^{i}}\frac{c}{n(\mathbf{r})}P^{k}\right) +%
\frac{\hbar ^{2}}{4}\frac{c}{n(\mathbf{r})}\left( \frac{1}{P}\partial
_{i}\ln n-\frac{P_{i}P_{j}}{P^{3}}\partial _{j}\ln n\right)
\]
Interestingly the corrective term does not give any contribution to the
velocity at this order of approximation so that the velocity is indeed given
by Eq. \ref{velo} at the order $\hbar ^{2}$

\[
v=\frac{c}{n(\mathbf{r})}\left( 1+\frac{\lambda ^{2}}{P^{2}}\left( \left(
\nabla \ln n\right) ^{2}-\frac{1}{P^{2}}\left( \mathbf{P.}\nabla \ln
n\right) ^{2}\right) \right) ^{1/2}+O(\hbar ^{3})
\]
It is worth mentioning that in the context of photons propagating in a
smoothly inhomogeneous medium we are apparently led to a contradiction with
known results. Indeed, it is known from optics that the second order
corrections in a smoothly inhomogeneous medium are responsible for the
linear birefringence, i.e., the equations of motion depend on the linear
polarization of the particle \cite{POLARIZATION} and not only on helicity as
found here. One possible explanation of this discrepancy might be that here
the photon in an inhomogeneous medium is treated as a quantum particle
moving in a static gravitational field, so that the medium index is
equivalent to a metric. This approach is perhaps too restrictive since by
doing so, we take only into account the geometry and neglect all
electrodynamical processes which are relevant for the interpretation of the
linear birefringence \cite{POLARIZATION}. In other words, an inhomogeneous
medium might be considered to be equivalent to a gravitational field at the
semiclassical order only, since the electrodynamics plays a role at the
second order. If true, our approach beyond the first order is strictly
speaking only valid for spinning particles in a static gravitational field.

\subsection{The electron in a time dependent electric field}

As a second example, we consider a spinless non-relativistic electron in a
periodic potential submitted to an external time dependent electric field.
We limit ourself to this simple case because a second order computation can
easily be performed as shown below, and leave more complicated cases with a
position dependent electric field and a magnetic field to future works.
Because of the non-trivial vacuum configuration, that is the band structure
of the energy spectrum in solid state physics this example fits in our
formalism similarly to the Dirac equation. Important new results where
recently found in the context of Bloch electron. Indeed, the semi-classical
equations of motion where found to be modified by Berry phase terms \cite
{NIU1}\cite{PIERREGENERAL}\cite{PIERREBLOCH}\cite{ADAM}, changing profoundly
the properties of electron transport in a solid \cite{NIULIOUVILLE}. It is
thus interesting to go beyond the semi-classical approximation to check
wether new effects might be revealed.

\subsubsection{First order}

Since the initial diagonalization matrix when $\alpha =0$ is not know, the
results obtained are more formal. Nevertheless, let us start with the $0$th
order formal diagonalization. Starting with the Hamiltonian of an electron
in a periodic potential plus an external perturbation
\begin{equation}
H_{0}(\mathbf{P},\mathbf{R})=\frac{\mathbf{P}^{2}}{2m}\mathbf{+}V(\mathbf{R)+%
}v(\mathbf{R)}
\end{equation}
with $v(\mathbf{R)=E.R}$ and $\mathbf{E(}t\mathbf{)}$ a time dependent
electric field, the zeroth order diagonalization can be done formally with a
$\mathbf{K}$ dependent unitary matrix $U_{\alpha }\left( \mathbf{K}\right) $%
, where $\mathbf{K}$ is the pseudo-momentum of the Bloch bands \cite
{PIERREBLOCH} \cite{PIERREGENERAL}. Therefore, only the position acquires a
Berry phase, i.e., $\mathcal{A}_{\alpha }^{R}\left( \mathbf{K}\right) =i%
\left[ U\nabla _{P}U^{+}\right] ^{+}$ and $\mathcal{A}_{\alpha }^{P}=-i\left[
U\nabla _{R}U^{+}\right] ^{+}=0+O\left( \alpha \right) $ implying $\mathcal{A%
}_{\alpha }^{P_{l},R_{l}}=0+O\left( \alpha \right) .$

As a consequence, the diagonalized Hamiltonian $\varepsilon _{\alpha }(%
\mathbf{K,R)}$ at order $\alpha $, projected on a particular band of index $%
n $ is given by :
\begin{equation}
\varepsilon _{\alpha ,n}(\mathbf{K,\mathbf{r}}_{n}\mathbf{)}=\widetilde{%
\varepsilon }_{n}(\mathbf{K)}+v\left( \mathbf{r}_{n}\right)
\end{equation}
where $\widetilde{\varepsilon }_{n}(\mathbf{K)}$ is the energy of the
unperturbed $n$th band (i.e. when $v(\mathbf{R)=}0)$ and $\mathbf{r}_{n}$
the covariant position operator defined as the projection of $\mathbf{%
R+\alpha }\mathcal{A}_{\alpha }^{R}$ on the $n$th band:
\begin{equation}
\mathbf{r}_{n}=\mathbf{R+\alpha }\mathcal{A}_{n,\alpha }^{R}\left( \mathbf{K}%
\right)
\end{equation}
where $\mathcal{A}_{\alpha ,n}^{R}$ is the projection of the matrix $%
\mathcal{A}_{\alpha }^{R}$ on the $n$th band.The equations of motion are
easily deduced to be
\begin{eqnarray}
\mathbf{\dot{r}}_{n} &=&\nabla _{\mathbf{K}}\varepsilon _{\alpha ,n}+\frac{1%
}{\hbar }\mathbf{\dot{K}}\mathbf{\times }\Theta _{n}^{rr}  \nonumber \\
\mathbf{\dot{K}} &=&\nabla _{\mathbf{r}}\varepsilon _{\alpha ,n}=\mathbf{E}
\label{eqmotionbloch}
\end{eqnarray}
with the Berry curvature defined as $\left[ r_{n,i},r_{n,j}\right] =\Theta
_{n,ij}^{rr}=\partial _{i}\mathcal{A}_{n,\alpha }^{R}-\partial _{j}\mathcal{A%
}_{n,\alpha }^{R}.$ The second term in Eq. \ref{eqmotionbloch} is called the
anomalous velocity and it is was shown todrive important new effects, such
as the discovery of a monopole in momentum contribution to the conductivity
of Bloch electrons in particular materials \cite{FANG}. It is then
interesting to check if this equation is changed beyond the semiclassical
level.

\subsubsection{Second Order and more}

The computation of the energy to the second order needs the value of the
transformation matrix at the first order in $\alpha .$ Consider therefore
the $0$th order differential equations for the transformation matrix Eq. \ref
{eq2}
\begin{equation}
0=\left[ \partial _{\alpha }U_{\alpha }\left( \mathbf{K},\mathbf{R}\right)
U_{\alpha }^{+}\left( \mathbf{K,R}\right) ,\varepsilon _{\alpha }\left(
\mathbf{K,R}\right) \right] ^{-}+\frac{1}{2}\left[ \mathcal{A}_{\alpha
}^{R_{l}}\nabla _{R_{l}}\varepsilon _{\alpha }\left( \mathbf{K,R}\right)
+\nabla _{R_{l}}\varepsilon _{\alpha }\left( \mathbf{K,R}\right) \mathcal{A}%
_{\alpha }^{R_{l}}\right] ^{-}  \label{eqelec1}
\end{equation}
and Eq. \ref{eq3}
\begin{equation}
0=\partial _{\alpha }U_{\alpha }(\mathbf{K},\mathbf{R)}U_{\alpha }^{+}(%
\mathbf{K,R)+}U\mathbf{_{\alpha }(K,\mathbf{R)}}\partial _{\alpha }U\mathbf{%
_{\alpha }^{+}(K\mathbf{,R)}}  \label{eqel2}
\end{equation}
We see that nothing ensures that : $\left[ \mathcal{A}_{\alpha
}^{R_{l}}\nabla _{R_{l}}\varepsilon _{\alpha }\left( \mathbf{K,R}\right)
+\nabla _{R_{l}}\varepsilon _{\alpha }\left( \mathbf{K,R}\right) \mathcal{A}%
_{\alpha }^{R_{l}}\right] ^{-}=0$. \ To solve for the transformation matrix,
we proceed as for the photon and decompose at the first order $U_{\alpha }(%
\mathbf{K},\mathbf{R)=}U_{0}(\mathbf{K},\mathbf{R)+}\alpha U_{1}(\mathbf{K},%
\mathbf{R)}$, so that, taking into account the specific form of \ $v(\mathbf{%
R)}$ our equations Eqs. \ref{eqelec1} and \ref{eqel2} reduce to
\begin{eqnarray*}
\left[ U_{1}(\mathbf{K},\mathbf{R)}U_{0}^{+}(\mathbf{K},\mathbf{R),}%
\varepsilon _{\alpha }\left( \mathbf{K,R}\right) \right] &=&-\left[ \mathcal{%
A}_{\alpha }^{R_{l}},E_{l}\right] ^{-} \\
U_{1}(\mathbf{K},\mathbf{R)}U_{0}^{+}(\mathbf{K},\mathbf{R)+}U_{1}(\mathbf{K}%
,\mathbf{R)}U_{0}^{+}(\mathbf{K},\mathbf{R)} &=&0
\end{eqnarray*}
Introducing the band indices, one easily sees that these two equations
reduce to the $2n^{2}-n$ following independent equations for $2n^{2}$ real
variables ($n$, the number of bands can of course be infinite; this does not
invalidate the argument)
\begin{eqnarray*}
\left[ U_{1}(\mathbf{K},\mathbf{R)}U_{0}^{+}(\mathbf{K},\mathbf{R)}\right]
_{mn}\left( \varepsilon _{\alpha }\left( \mathbf{K,R}\right)
_{n}-\varepsilon _{\alpha }\left( \mathbf{K,R}\right) _{m}\right) &=&-\left[
\mathcal{A}_{\alpha }^{R_{l}},E_{l}\right] _{mn}\text{, for }m<n \\
\left[ U_{1}(\mathbf{K},\mathbf{R)}U_{0}^{+}(\mathbf{K},\mathbf{R)}\right]
_{mn}+\left[ U_{1}(\mathbf{K},\mathbf{R)}U_{0}^{+}(\mathbf{K},\mathbf{R)}%
\right] _{mn} &=&0\text{ }
\end{eqnarray*}
Once again, we are left with a gauge choice on the diagonal. We can simply
impose as a condition that the diagonal elements are null. Actually, our
equation imposes that $\left[ U_{1}(\mathbf{K},\mathbf{R)}U_{0}^{+}(\mathbf{K%
},\mathbf{R)}\right] _{nn}$ is an imaginary number. As a consequence,
similarly to the photon, composing $U_{\alpha }(\mathbf{K},\mathbf{R)}$ on
the left with the unitary matrix $X$ defined by
\begin{equation}
X_{mn}=1-\alpha \delta _{mn}\left[ U_{1}(\mathbf{K},\mathbf{R)}U_{0}^{+}(%
\mathbf{K},\mathbf{R)}\right] _{nn}
\end{equation}
yields a diagonalization matrix with zero on the diagonal.

Having chosen the gauge, the equations for $U_{1}(\mathbf{K},\mathbf{R)}%
U_{0}^{+}(\mathbf{K},\mathbf{R)}$ can be solved formally, through the Cramer
rules. Fortunately, given our very simple case of potential, we get
\begin{equation}
\varepsilon _{\alpha }\left( \mathbf{K,R}\right) _{n}-\varepsilon _{\alpha
}\left( \mathbf{K,R}\right) _{m}=\tilde{\varepsilon}_{\alpha }\left( \mathbf{%
K}\right) _{n}-\tilde{\varepsilon}_{\alpha }\left( \mathbf{K}\right) _{m}
\end{equation}
which does not depend on $\mathbf{R.}$ As a consequence, $U_{1}(\mathbf{K},%
\mathbf{R)}U_{0}^{+}(\mathbf{K},\mathbf{R)}$ and thus $U_{1}(\mathbf{K},%
\mathbf{R)}$ is only a function of $\mathbf{K}$.

We can thus write
\begin{equation}
U_{\alpha }(\mathbf{K},\mathbf{R)=}U_{0}(\mathbf{K)+}\alpha U_{1}(\mathbf{K)}
\end{equation}
As a consequence at the first order we will have again.:
\begin{equation}
\mathcal{A}_{\alpha }^{R}=i\left[ U_{\alpha }\nabla _{K}U_{\alpha }^{+}%
\right] ^{+}+O\left( \alpha ^{2}\right)
\end{equation}
and for the other Berry curvatures we have $\mathcal{A}_{\alpha
}^{P}=0+O\left( \alpha ^{2}\right) ,$ and $\mathcal{A}_{\alpha
}^{P_{l},R_{l}}=0+O\left( \alpha ^{2}\right) .$ Let us just mention that the
Berry phase in $R$ is now different from its the zeroth order value. It has
now a term proportional to $\alpha $ which is included in its definition.

Consider now the differential equation for the energy at the first order in $%
\alpha $. A simplification arise, since here, at the first order \bigskip
\begin{equation}
Asym\left\{ \nabla _{P_{l}}\nabla _{R_{l}}\varepsilon _{\alpha }\left(
\mathbf{K,R}\right) \right\} -U_{\alpha }Asym\left\{ \nabla _{P_{l}}\nabla
_{R_{l}}H_{0}\left( \mathbf{P,R}\right) \right\} U_{\alpha }^{+}=0
\end{equation}
actually $\mathbf{K}$ and $\mathbf{R}$ are not mixed in $\varepsilon
_{\alpha }$ or $H_{0}$. As a consequence, and given our previous results,
the projection of our equation on the diagonal part leads to
\begin{equation}
\frac{d}{d\alpha }\varepsilon _{\alpha }\left( \mathbf{K,R}\right) =\frac{1}{%
2}\left[ \mathcal{A}_{\alpha }^{R_{l}}\nabla _{R_{l}}\varepsilon _{\alpha
}\left( \mathbf{K,R}\right) +\nabla _{R_{l}}\varepsilon _{\alpha }\left(
\mathbf{K,R}\right) \mathcal{A}_{\alpha }^{R_{l}}\right] ^{+}
\end{equation}
Exactly as in the photon case, this equation has the second order solution
after projection on the $n$th band
\begin{equation}
\varepsilon _{\alpha ,n}\left( \mathbf{K,r}_{n}\right) =\tilde{\varepsilon}%
_{n}(\mathbf{K)}+v\left( \mathbf{r}_{n}\right)
\end{equation}
where now $\mathbf{r}_{n}=\mathbf{R+}\alpha \mathcal{A}_{\alpha ,n}^{R}$.
Note that $\alpha \mathcal{A}_{\alpha ,n}^{R}$ is actually of order $\alpha
^{2}$. Therefore the energy has formally the same expression than in the
first order case. The only difference stems from the potential energy
because the position operator get a $\alpha ^{2}$ contribution through the
presence of the Berry connection.

As a consequence, the equations of motion are formally the same as for the
first order (removing the band index), i.e.,
\begin{eqnarray*}
\mathbf{\dot{r}}_{n} &=&\nabla _{\mathbf{K}}\varepsilon _{\alpha ,n}+\frac{1%
}{\hbar }\stackrel{\bullet }{\mathbf{K}}\mathbf{\times }\Theta _{n}^{rr} \\
\stackrel{\bullet }{\mathbf{K}} &=&\nabla _{\mathbf{r}_{n}}\varepsilon
_{\alpha ,n}=\mathbf{E}
\end{eqnarray*}
The only difference with respect to the semiclassical equations of motion Eq.%
\ref{eqmotionbloch} is that the Berry connection gets a contribution of
order $\hbar ^{2}$ but formally the equations of motion are unchanged.

To conclude this section let us state the following important result. One
can easily check that all our construction can be generalized to all orders
in $\hbar $. As a consequence, the Hamiltonian form for $\varepsilon
_{\alpha ,n}\left( \mathbf{K,r}_{n}\right) $ is valid to all orders, and the
equation of motions are in fact quantum operator equations. The only
difference appearing when diagonalizing recursively, consists in new
contributions to the Berry connection.

This is a important new result, because these equations were considered
until now has semiclassical equation of motion with the usual restrictions
imposed in this case. It would be interesting to check if this result is
generalizable to spatially variable electric field and in the presence of a
magnetic field, but then the equations get quickly cumbersome.

\section{CONCLUSION}

Some recent applications of semi classical methods to several branches of
Physics, such as spintronics or solid state physics have shown the relevance
of Berry Phases contributions to the dynamics of a system, leading notably
to the discovery of the intrinsic spin Hall effect. However, these
progresses called for a rigorous Hamiltonian treatment that would allow for
deriving naturally the role of the Berry phase in theses systems and to go
beyond the semiclassical level. In this paper, we presented a new
diagonalization method for a generic matrix valued Hamiltonian, based on a
formal expansion in power of $\hbar $. A differential equation connecting
two diagonalization processes for two very close values of $\hbar $,
considered as a running parameter, was derived. The integration of this
differential equation allows the recursive determination of the series
expansion of the diagonalized Hamiltonian in powers of $\hbar $. This
approach, which results in effective Hamiltonians with Berry phase
corrections, goes beyond previous works on the semiclassical diagonalization
of quantum Hamiltonians. The resulting generic equations of motion are also
corrected by Berry phase terms of higher order in $\hbar $. As physical
applications, we considered a spinning massless particle propagating in a
isotropic inhomogeneous medium and showed that both the energy and the
velocity get quantum corrections of order $\hbar ^{2}$. We also derived
formally to all order in $\hbar $ the energy spectrum and the equations of
motion of Bloch electrons in an external constant electric field, showing
that the equations of motion are actually quantum operator equations (valid
to each order in $\hbar $). It would be interesting to check if this result
persists for variable electric fields and in the presence of magnetic
fields. Our approach is a general one and will be applied in the future to
other condensed matter systems as well as in particle physics. \bigskip

\textbf{Acknowledgment.} The authors acknowledge a fruitful correspondence
with K. Bliokh about the linear birefringence phenomenon discussed in the
section IV.2., as well as for having brought to our knowledge reference \cite
{POLARIZATION}

\end{document}